\documentstyle[12pt]{article}  
\newcommand{\be}{ \begin{equation}}
\newcommand{\ee}{\end{equation}} 
\begin{document} 
\def\theequation{\arabic{section}.\arabic{equation}} 
\begin{titlepage} 
\title{de Sitter attractors in generalized gravity} 
\author{Valerio Faraoni\\ \\ 
{\small \it Physics Department, University of Northern 
British Columbia}\\ 
{\small \it 3333 University Way, Prince George, B.C., Canada V2N~4Z9}\\ 
{\small \it  email~~vfaraoni@unbc.ca} }
\date{} \maketitle 
\thispagestyle{empty} 
\vspace*{1truecm} 
\begin{abstract} 
We obtain conditions for the existence and stability of de Sitter attractors in the phase 
space of homogeneous and isotropic cosmology in generalized theories of
gravity (including non-linear and scalar-tensor theories). These conditions are valid  
for any form of the coupling functions of the 
theory. Stability  with respect to inhomogeneous perturbations is analyzed using a 
covariant and gauge-invariant formalism. The relevance for inflationary scenarios of the early 
universe and for quintessence models of the present era is discussed.  
\end{abstract} 
\vspace*{1truecm} 
%\noindent {\bf Keywords:}
%generalized gravity, scalar-tensor theories, de Sitter space, inflation, quintessence.  
%\begin{center}  Submitted for publication \end{center}
\end{titlepage}

\def\theequation{\arabic{section}.\arabic{equation}}

%%%%%%%%%%%%%%%%%%%%%%%%%%%%%%%%%%%%%%%%%%%%%%%%%%%%%%%%%%%%%%%%%%%%%%%%%%%%%%%%%%%%

\section{Introduction}
\setcounter{equation}{0}

In general relativity de Sitter space plays a special role because quantum field theory predicts 
the existence of vacuum energy, which is equivalent to a cosmological constant $\Lambda$, and the 
solution of the Einstein field equations with vacuum energy as the only material source   
is de Sitter  space. A period of de Sitter-like inflationary expansion of the early universe 
has come to be
regarded as the canonical solution to the horizon, flatness and monopole
problems that plague standard big bang cosmology. As a bonus, 
inflation provides a mechanism for generating density perturbations through quantum 
fluctuations of the inflaton field, seeding the structures observed in the universe today
\cite{KT,Lidseyetal,MukhanovFeldmanBrandenberger}.

In most inflationary models based on general relativity the expansion of the universe described 
by the scale factor 
$a(t)$ of the Friedmann-Lemaitre-Robertson-Walker (hereafter ``FLRW'') metric 
\be   \label{I1}
ds^2=-dt^2 +a^2 (t) \left( dx^2 + dy^2 + dz^2 \right)  \; ,
\ee
is approximately exponential. 
This accelerated expansion is achieved if the dynamics of the universe are 
dominated by a scalar 
field $\phi$  self-interacting through a potential $V( \phi ) $ that has a plateau 
such that $V( \phi ) \simeq V_0 =$constant for a certain range of values of $\phi$. While $\phi $ 
evolves through 
this interval the potential mimics a cosmological constant. The 
corresponding solution of the Einstein equations has the form 
\be \label{I2}
a(t)=a_0 \, \mbox{e}^{H(t) \, t} 
\ee
where 
\be \label{I3}
H(t)=H_0+H_1\, t + \, ...
\ee
and $a_0, H_0$, and $H_1$ are constant, with $ \left| H_1 t \right| << H_0 $.  
In other words spacetime is close to the de Sitter solution and the scalar field rolls slowly over 
the plateau of the potential (``slow-roll approximation'' \cite{LiddleLyth00,Lidseyetal}). 
The dynamics of a scalar field  minimally coupled to 
the spacetime curvature are described by the Klein-Gordon 
equation
\be \label{KG}
\ddot{\phi}+3H \, \dot{\phi}+\frac{ dV}{d\phi}=0 \;.
\ee
The flat section of  the 
potential does not guarantee that the solution of the field equations is of the form 
(\ref{I2}) and (\ref{I3}). The fact that the scalar $\phi(t)$ rolls slowly corresponds to neglecting 
its first derivative (its ``speed'') $\dot{\phi}$ in the Klein-Gordon equation (\ref{KG}), which 
reduces to $ \ddot{\phi} \simeq -\,  dV / d\phi $.
Alternatively, the slow-roll approximation corresponds to neglecting the kinetic energy density  $ 
\left( \dot{\phi} \right)^2/2$ in the expressions of  the scalar field energy density and pressure
\be  \label{I6}
\rho =  \frac{\dot{\phi}^2}{2} -V ( \phi )
\; ,
\ee
\be  \label{I7}
P= \frac{\dot{\phi}^2}{2} + V ( \phi ) \; .
\ee
As a result, the scalar field is equivalent to a fluid with equation of state $P \simeq - \rho$. 
For comparison, in de Sitter space the cosmological 
constant $\Lambda$ can be regarded as a matter fluid with energy density and pressure
\be  \label{I8}
\rho_{\Lambda} = \frac{\Lambda}{8\pi G} \; , \;\;\;\;\;\;
P_{\Lambda} = -\, \frac{\Lambda}{8\pi G} \; , 
\ee
and equation of state $P_{\Lambda}=-\rho_{\Lambda}$.

Even if the potential $V( \phi ) $ has a flat section, the scalar $\phi$ could still roll fast  (with 
non-negligible $\dot{\phi}$) over it -- the slow-roll approximation is an assumption about the 
solution 
$\left( a(t), \phi(t) \right)$ of the dynamical equations, not on the form of $V( \phi )$. What 
makes this approximation viable is 
the fact that, in general relativity with a minimally coupled scalar field (and also when the field 
is non-minimally coupled \cite{GunzigetalCQG,FaraoniPLA,FaraoniIJTP}), de Sitter space is an 
attractor  for the orbits of the solutions in phase space \cite{LiddleLyth00}. The main purpose 
of the present paper is to establish whether de Sitter space is an attractor also in more 
general gravity theories.

There has been increasing interest in cosmology in alternative theories of gravity, with 
several different  motivations \cite{VF04}. One such motivation arises in the quest for a 
quantum theory 
of gravity: it is widely believed 
that quantum  corrections modify the Einstein-Hilbert gravitational Lagrangian by adding terms 
proportional to higher order curvature invariants \cite{Stelle78}.  
These  corrections to the classical Lagrangian are small at 
small curvatures but become dominant when $R$ grows, e.g., approaching a singularity.

From another point of view, theories of gravity generalizing Einstein's 
relativity have been studied for decades at the classical level \cite{VF04}. The 
prototypical 
alternative theory, Brans-Dicke theory, was 
originally motivated by the need to explicitly incorporate  Mach's principle in relativistic 
cosmology, and has later been generalized to the class of scalar-tensor theories in which 
a Brans-Dicke-like scalar describes the gravitational field together with the metric 
tensor, and 
coupling functions appear in scalar-tensor gravity instead of coupling 
constants.  In versions of these theories motivated by high energy physics the Brans-Dicke-like 
scalar $\phi$ is allowed to self-interact through a potential $V( \phi)$. The 
gravitational sector of scalar-tensor theories is described by the action
\be\label{I9}
S_{ST} = \int d^4 x \, \sqrt{-g} \left[ \frac{ f( \phi )}{2} \, R -\frac{\omega( \phi )}{2}\, g^{ab} 
\nabla_a \, \phi \, 
\nabla_b \,\phi -V( \phi) \right] \;.
\ee
Scalar-tensor gravity has been studied in relation to inflation, resulting in various extended 
\cite{extended} and hyperextended \cite{hyperextended} inflationary scenarios. Added interest comes 
from the fact that a gravitational scalar field is an essential ingredient of modern high 
energy theories unifying gravity with the other fundamental interactions (in particular string 
theories) \cite{VF04}, from certain similarities between scalar-tensor and string theories 
\cite{VF04}, and from the fact 
that  the low-energy limit of the bosonic string theory is a Brans-Dicke theory with  
parameter  $\omega=-1$ \cite{string}. 

In this paper  non-linear gravity and scalar-tensor theories are considered simultaneously as 
special cases 
of the generalized gravity theory described by the action
\be \label{I10}
S = \int d^4 x \, \sqrt{-g} \left[ \frac{1}{2} \, f( \phi, R )  -\frac{1}{2} \, \omega( \phi ) 
g^{ab} \nabla_a \, \phi \,  \nabla_b \,\phi -V( \phi) \right] \;.
\ee
We neglect matter contributions to the action because we want to study situations in which the scalar 
$\phi$ dominates the dynamics of the universe, such as during inflation in the early universe or in a 
late era in which a quintessence scalar field has come to dominate.
The action contains the Ricci scalar $R$ but no other curvature invariant. In addition to the 
simplification that accompanies it, this choice is motivated by the fact that in the homogeneous and 
isotropic 
cosmologies that we consider all the quadratic invariants of the Riemann tensor can be 
expressed in terms of $R$. The action (\ref{I10}) includes as special cases 
Brans-Dicke theory, scalar-tensor theories, induced gravity, quadratic Lagrangians, the theory 
of a scalar field coupled non-minimally to the the Ricci scalar, general relativity with or 
without  a minimally coupled scalar field and  a cosmological constant, and the theory of 
phantom fields.

In the literature, slow-roll inflation in the context of generalized gravity is often considered. 
It is interesting to determine whether de Sitter spaces are 
actually  solutions of the theory and whether they are attractors for the orbits of the solutions. 
This issue is crucial for understanding inflation in generalized gravity because the slow-roll 
approximation is meaningless unless there  is a de Sitter attractor in phase space.

Another independent motivation comes from the recent 
discovery \cite{SN} that the present expansion of the universe is accelerated, which has led 
cosmologists to  postulate the existence of a new form of of energy called {\em quintessence} or 
{\em dark  energy} with 
the exotic equation of state $P< -\rho/3$. Indeed, there are claims of evidence for a very negative
 pressure $P< -\rho$, a fact that, if confirmed, has interesting implications for the future of the 
universe -- it could lead to a 
Big Rip singularity in a finite future \cite{bigrip} (see Sec.~6 for a discussion). An obvious 
candidate 
for dark energy is the cosmological constant associated with de Sitter space. However, the 
cosmological 
constant carries with it two embarassing problems: 1) the well-known cosmological constant problem 
\cite{Weinberg89} of why the value of $\rho_{\Lambda} $ predicted by quantum field theory is 120 
orders of magnitude larger than the energy density of the universe; and 2) the cosmic coincidence 
problem of why the dark energy is 
beginning to dominate the cosmic dynamics right now when there are galaxies and human observers to 
notice it. These  problems are only solved by an enormous amount of fine-tuning.  For these reasons, 
theoretical models of quintessence explore different avenues. Among the many models proposed, 
modifications of  Einstein 
gravity including non-linear corrections to the Einstein-Hilbert action have been proposed, in both the 
Einstein-Hilbert 
\cite{Carrolletal03}-\cite{SoussaWoodard03}
and the Palatini form of the variational 
principle \cite{Vollick}-\cite{Flanagan}. Such models do not usually 
admit a Minkowski solution that would be useful to study the weak-field limit of the theory -- 
a de Sitter space is used 
instead for this purpose. Moreover,  quintessence models that do not end in a Big Rip  often 
evolve to a de Sitter phase in the future.   Thus, both classes of models -- either  invoking 
a scalar field as dark energy (in general relativity or in scalar-tensor gravity), or 
advocating non-linear corrections to gravity, exhibit aspects related to the existence of de 
Sitter solutions.

 In a different context, it is interesting to 
examine the stability of general relativity with respect to small deviations 
from Einstein's theory due to quantum corrections. This is the approach adopted, e.g., in 
Ref.~\cite{BarrowOttewill83}.

The purpose of the present paper is to establish conditions under 
which de Sitter solutions exist in the generalized theory described by the action (\ref{I10}), and 
to study  their stability with respect to inhomogeneous perturbations. Our main motivation is 
to  
establish a  firm foundation for the slow-roll approximation to de Sitter-like inflation in 
these theories.

The issues of existence and  stability of de Sitter solutions have been addressed in the literature 
only for special cases of the general theory (\ref{I10}) and usually only for spatially homogeneous 
perturbations. This limitation is probably due to the fact that inhomogeneous perturbations 
are in general gauge-dependent and they must be analyzed in the context of a covariant and 
gauge-invariant 
formalism. The latter is 
substantially more complicated than the analysis of time-dependent homogeneous 
perturbations. In the present paper the covariant and gauge-invariant formalism of 
Bardeen-Ellis-Bruni-Hwang-Vishniac is employed to study stability. This formalism has been used 
before to analyze the stability of de Sitter solutions in the theory of a scalar field coupled 
non-minimally to the curvature \cite{FaraoniPLA,FaraoniIJTP}, and the stability of Einstein 
space in 
general relativity \cite{Barrowetal03}. Following the same line of reasoning, it is also interesting 
to  consider the stability of Minkowski space solutions of 
the theory.

An independent motivation for the study of de Sitter space arises from the idea that the universe 
could have originated in a de Sitter state, thus avoiding the initial big bang singularity and 
evolving into an inflationary phase. Variations of this idea include the possibility of a Minkowski 
\cite{Wesson85}-\cite{Fabrisetal02} or an Einstein space 
\cite{EllisMaartens02,Barrowetal03} as a possible initial state. We include Minkowski space in 
our analysis as a special case of de Sitter space.

The plan of this paper is as follows. In Sec.~2 we summarize the field equations of the generalized 
theory and we derive the conditions for the existence of de Sitter solutions. Section~3 addresses the 
issue of stability with respect to inhomogeneous perturbations using a covariant and 
gauge-invariant approach. 
Section~4 discusses the existence and stability of Minkowskian solutions of generalized gravity, 
while 
Sec.~5 contains a discussion and the conclusions. 
We use units in which the speed of  light $c=1$ and $8\pi G=1$, where $G$ is Newton's 
constant, the metric signature is $-,+,+,+ $, and  $\Box \equiv g^{ab}\nabla_a \nabla_b$ 
denotes d'Alembert's operator. For ease of comparison with previous works, the other 
conventions follow Refs.~\cite{Hwang,HwangCQG}.

%%%%%%%%%%%%%%%%%%%%%%%%%%%%%%%%%%%%%%%%%%%%%%%%%%%%%%%%%%%%%%%%%%%%%%%%%%

\section{Generalized gravity, fixed points, and de Sitter solutions}
\setcounter{equation}{0}

Variation of the action (\ref{I10}) leads to the field equations of generalized gravity 
\cite{HwangCQG}
\begin{eqnarray}
 G_{ab} & = & \frac{1}{F} \left[ \omega \left( \nabla_a \, \phi \nabla_b \, \phi - \, 
\frac{1}{2} \,  g_{ab} 
\nabla^c \, \phi \nabla_c \, \phi \right) -\frac{1}{2} \, g_{ab} \left( RF-f+2V \right) 
\right. \nonumber \\
&& \nonumber \\
& & \left. +  \nabla_a \nabla_b \, F -g_{ab} \Box F \right] \;,  \label{2}
\end{eqnarray}

\be \label{3}
\Box \phi +\frac{1}{2\omega} \left( \frac{d\omega}{d\phi} \, \nabla^c \, \phi \nabla_c \, \phi
+\frac{\partial f}{\partial \phi} -2\, \frac{dV}{d\phi} \right)=0 \;,
\ee
where 
\be \label{4}
F \equiv \frac{\partial f}{\partial R} \;.
\ee
For a FLRW metric of curvature index $K$, given by the line element
\be   \label{4bis}
ds^2=-dt^2 +a^2 (t) \left[ \frac{dr^2}{1-Kr^2} + r^2 \left(d\theta^2 + \sin^2 \theta \,\, d\varphi^2 
\right) \right] 
\ee
in comoving coordinates $\left( t,r, \theta, \varphi \right)$, the field equations assume the form
\begin{eqnarray} 
H^2 & =&  \frac{1}{3F} \left( \frac{\omega}{2} \, \dot{\phi}^2 +\frac{RF}{2} -\frac{f}{2} +V 
-3\dot{H}F 
\right) -\frac{K}{a^2} \;, \label{5} \\
&& \nonumber \\
\dot{H} & = &  - \, \frac{1}{2F} \left( \omega \dot{\phi}^2 + \ddot{F}  -H\dot{F} \right) 
+\frac{K}{a^2} \;,
 \label{6} 
\end{eqnarray}

\be \label{7}
\ddot{\phi } +3 H \dot{\phi} +\frac{1}{2\omega} \left(
\frac{d\omega}{d\phi} \,  \dot{\phi}^2 - \frac{\partial f}{\partial \phi} +2\, \frac{dV}{d\phi}  
\right) =0 \;,
\ee
where $H\equiv \dot{a}/a$ is the Hubble parameter, an overdot denotes differentiation with respect to 
the comoving time $t$ and the Ricci curvature is $R= 6\left( \dot{H}+2H^2 +K/a^2 \right)$. Only two 
equations in the set (\ref{5})-(\ref{7}) are independent.

There is now substantial evidence that the universe has flat 
spatial sections \cite{Boomerang} and therefore from now on we restrict ourselves to the spatially 
flat case $K=0$. In this case one can choose $ H $ and $\phi$ as dynamical variables -- this is not 
possible if $K \neq 0$, in which case one  has to consider as dynamical variable the 
scale  factor $a$ appearing in the field equations through the terms $\pm K/a^2$ instead of $H$. 
However when $K=0$ these terms disappear and $a$ appears only in the combination $ H=\dot{a}/a$ and 
in its time derivatives (in non-linear gravity the field equations are of fourth order  and 
$ \ddot{H} $ and $ \stackrel{ \cdot\cdot\cdot}{H} $ appear in the field equations).  The phase 
space 
picture of the 
dynamical system depends on the specific form of the functions $f( \phi, R ) $, $\omega 
( \phi ) $ and $V( \phi )$. However, for any choice of $f, \omega$ 
and $V$, the fixed points of the system 
(if they exist) are given by
\be \label{desitter}
\left( H, \dot{H}, \ddot{H},\stackrel{ \cdot\cdot\cdot}{H} , \phi , \dot{\phi}  \right) 
=\left( H_0, 0,0, 0, \phi_0 , 0 \right) 
\;,
\ee
where $H_0$ and $\phi_0$ are constants, i.e., they are de Sitter spaces with constant scalar field 
\cite{footnote2}. The conditions for the existence of de Sitter fixed point solutions are obtained by 
substituting eq.~(\ref{desitter}) in eqs.~(\ref{5}) and (\ref{7}), which yields the two conditions
\be \label{14}
6H_0^2 \, F_0 - f_0+2V_0 =0 
\ee
and
\be
{f_0}'- 2{V_0}' =0 \;, \label{15} 
\ee
where 
\begin{eqnarray} 
F_0 & \equiv & \left. \frac{\partial f}{\partial R} \right|_{ \left( \phi_0, R_0 \right)} \;, 
\\
&& \nonumber \\
f_0 & \equiv &  f \left( \phi_0, R_0 \right) \;,  \\
&& \nonumber \\
V_0 & \equiv & V  \left( \phi_0 \right) \;,  \\
&& \nonumber \\
{V_0}' & \equiv & \left. \frac{dV}{d\phi}  \right|_{ \phi_0 } \;,  \\
&& \nonumber \\
{f_0}' & \equiv & \left. \frac{\partial f}{\partial \phi} \right|_{\left( \phi_0, R_0 \right)} \;,
\end{eqnarray}
and $ R_0 = 12 H_0^2 $. There are two independent conditions (\ref{14}) and (\ref{15}) for the existence of de Sitter solutions 
because only two equations in the set (\ref{5})-(\ref{7}) are independent. 

Let us consider a few examples of specific gravity theories. Eq.~(\ref{14}) generalizes the 
condition 
\be
6H_0^2 F_0 -f_0 +2\Lambda =0 
\ee
found in Ref.~\cite{BarrowOttewill83} for the non-linear gravity 
theories given by the choice
\be
\phi=1 \;, \;\;\;\;\;\;\; f=f(R) \;, \;\;\;\;\;\; V=\Lambda=  \mbox{const.} \
\ee
(there is only one condition in this case because the scalar field is not a dynamical 
variable). 
Note that not all generalized gravity theories admit de Sitter solutions. For example, non-linear 
theories with  $ f( R ) = AR^n $,  $A=$constant, $n>2$ and $\phi=0$, $V=0$ do not 
satisfy eq.~(\ref{14}) \cite{BarrowOttewill83}.

In general relativity with a cosmological constant $\Lambda >0$ and without scalar eq.~(\ref{14}) 
produces the familiar de Sitter solutions 
\be
\left( H_0 , \phi_0 \right) = \left( \pm \sqrt{ \frac{ \Lambda}{3} }, 0 
\right) \;.
\ee
If a minimally coupled scalar is present, de Sitter space is achieved if 
\be V_0 > 0 \; ,\;\;\;\;\;\;\;\;  H_0 =\pm  
\sqrt{ \frac{V_0}{3} } \;,\;\;\;\;\; \mbox{and} \;\;\;\;\;\;\; V_0' =0 \;.
\ee

In the theory of a non-minimally coupled scalar field 
corresponding to 
\be \label{NMC}
f( \phi, R) = R\left( 1-\xi \phi^2 \right)  \;, \;\;\;\;\;\;\; \omega=1 \;, 
\ee
where $\xi $ is a dimensionless coupling constant, eqs.~(\ref{14}) and (\ref{15}) reduce to the 
conditions for the existence of de Sitter fixed points previously found in Refs. 
\cite{GunzigetalCQG}-\cite{FaraoniIJTP}
\be
H_0^2 \left( 1-\xi  \phi_0^2 \right) =\frac{V_0}{3}  \;,
\ee
\be
12\xi H_0^2 \, \phi_0 +V_0'=0 \;.
\ee

Recently the higher derivative theory of gravity described by 
\be  \label{questa}
f \left( \phi, R \right) = R-\,
\frac{\mu^2}{R} \;, \;\;\;\;\;\; \phi \equiv 1 \;, \;\;\;\;\; \omega=0 \; , \;\;\;\; V=0 \;,
\ee
where $\mu^{-1/2}$ is a length scale, has attracted attention as a model theory for the acceleration 
of the universe that does not require dark energy 
\cite{Carrolletal03}-\cite{Flanagan}. In this theory the correction to the 
Einstein-Hilbert Lagrangian is small at large curvatures, but becomes important as the universe 
expands and $R \rightarrow 0$. The conditions (\ref{14}) and 
(\ref{15}) for the existence of de Sitter solutions reduce to
\be
H_0 =\pm \frac{1}{2} \, \sqrt{ \frac{\mu}{ \sqrt{3}} } \;.
\ee
Note that this theory does not admit a Minkowski space solution corresponding to $H \equiv 0$.

%%%%%%%%%%%%%%%%%%%%%%%%%%%%%%%%%%%%%%%%%%%%%%%%%%%%%%%%%%%%%%%%%%%%%%%%%%%%%%%%%%%%

\section{Perturbations}
\setcounter{equation}{0}

If de Sitter fixed points exist for the dynamical system (\ref{5})-(\ref{7}) with $K=0$, the 
problem arises whether these fixed points are attractors in phase space or are unstable -- a 
stability analysis is required to answer this question. The approaches to this problem 
available in the literature 
\cite{Eddington30}-\cite{Gibbons} \cite{BarrowOttewill83,FaraoniPLA,FaraoniIJTP,
Fabrisetal00,Barrowetal03} 
are limited to 
special cases of generalized gravity theories and, usually, to homogeneous perturbations. The 
consideration of more general inhomogeneous perturbations is complicated by the 
gauge-dependence of this kind of cosmological perturbations. A gauge-independent 
analysis requires the use of a covariant and gauge-invariant formalism, which has been used before to 
study the  stability of de Sitter solutions  against inhomogeneous perturbations in the special case 
of the theory  described by  eq.~(\ref{NMC}) \cite{FaraoniPLA,FaraoniIJTP}. Another problem 
addressed in the literature with a gauge-independent approach is the 
stability of  the Einstein universe in general relativity with a non-minimally coupled scalar 
field 
\cite{Barrowetal03}.

We proceed by using the covariant and gauge-invariant formalism of Bardeen 
\cite{Bardeen,MukhanovFeldmanBrandenberger} further  
developed by Ellis, Bruni, Hwang and Vishniac \cite{EllisBruni,HwangVishniac}. A 
version for 
generalized theories of gravity is given in Refs.~\cite{HwangCQG,Hwang}. The metric perturbations 
are defined by
\begin{eqnarray} 
g_{00} & = & -a^2 \left( 1+2AY \right) \;, \label{19} \\
&& \nonumber \\
g_{0i} & = & -a^2 \, B \, Y_i  \;, \label{20} \\
&& \nonumber \\
g_{ij} & =& a^2 \left[ h_{ij}\left(  1+2H_L \right) +2H_T \, Y_{ij}  \right] \;,\label{21}
\end{eqnarray}
where the scalar harmonics $Y$ are the eigenfunctions of the eigenvalue problem
\be \label{22}
\bar{\nabla_i}
\bar{\nabla^i} \, Y =-k^2 \, Y \;.
\ee
Here $h_{ij} $ is the three-dimensional metric of the FLRW background and the operator $ \bar{\nabla_i} 
$ is the covariant derivative associated with  $h_{ij}$, while $k$ is an eigenvalue. The vector and tensor 
harmonics $Y_i$ and $Y_{ij}$ are defined by
\be \label{23}
Y_i= -\frac{1}{k} \, \bar{\nabla_i} Y \;,
\ee
\be \label{24}
Y_{ij}= \frac{1}{k^2} \, \bar{\nabla_i}\bar{\nabla_j} Y +\frac{1}{3} \, Y \, h_{ij} \;.
\ee
We use Bardeen's  \cite{Bardeen} gauge-invariant potentials $\Phi_H $ and $\Phi_A$ and the 
Ellis-Bruni \cite{EllisBruni} variable $\Delta \Phi $ defined by
\be \label{25}
\Phi_H = H_L +\frac{H_T}{3} +\frac{ \dot{a} }{k} \left( B-\frac{a}{k} \, \dot{H}_T \right) \;, 
\ee

\be \label{26}
\Phi_A = A  +\frac{ \dot{a} }{k} \left( B-\frac{a}{k} \, \dot{H}_T \right)
+\frac{a}{k} \left[ \dot{B} -\frac{1}{k} \left( a \dot{H}_T \right)\dot{}  \right] \;, 
\ee

\be \label{27}
\Delta \phi = \delta \phi  +\frac{a}{k} \, \dot{\phi}  \left( B-\frac{a}{k} \, \dot{H}_T 
\right) 
\;.
\ee
Equations analogous to eq.~(\ref{27}) define the gauge-independent variables $\Delta F, \Delta 
f$, and 
$\Delta R$. To first order the perturbations evolve according to the equations  \cite{HwangCQG}
\begin{eqnarray}  \label{29}
&& \Delta \ddot{\phi} + \left( 3H + \frac{\dot{\phi}}{ \omega} \, 
\frac{d\omega}{d\phi} \right) \Delta 
\dot{\phi} + \left[ \frac{k^2}{a^2}
 + \frac{\dot{\phi}^2}{2} \frac{d}{d\phi} \left( \frac{1}{\omega} \frac{d\omega}{d\phi} \right) -\, 
\frac{d}{d\phi} \left( \frac{1}{2\omega} \frac{\partial f}{\partial \phi} -\frac{1}{\omega} 
\frac{dV}{d\phi} \right) \right] \Delta \phi  \nonumber \\
&& \nonumber \\
&& =
 \dot{\phi}  \left(  \dot{\Phi}_A - 3\dot{\Phi}_H \right) 
+ \frac{\Phi_A}{\omega}  \left( \frac{\partial f}{\partial \phi} -2 \, \frac{dV}{d\phi} \right)  
+\frac{1}{2\omega} \, \frac{\partial^2 f}{\partial \phi \partial  R}  \, \Delta R  \; ,
\end{eqnarray}

\begin{eqnarray}  \label{30}
&& \Delta \ddot{F} +3H \Delta \dot{F} +\left( \frac{k^2}{a^2} - \frac{R}{3} \right) \Delta F 
+\frac{F}{3} \, \Delta R +\frac{2}{3} \, \omega \dot{\phi} \Delta \dot{\phi} 
+\frac{1}{3} \left( \dot{\phi}^2 \frac{d\omega}{d\phi} + 2\frac{\partial f}{\partial \phi} 
-4 \, \frac{dV}{d\phi} \right) \Delta \phi \nonumber \\
&& \nonumber \\
&& = \dot{F}  \left(  \dot{\Phi}_A - 3\dot{\Phi}_H \right) 
+ \frac{2}{3}  \left( FR -2f +4V  \right)  \Phi_A \; ,
\end{eqnarray}

\be \label{30bis}
\ddot{H}_T +\left( 3H+ \frac{\dot{F}}{F} \right) \dot{H}_T +\frac{k^2}{a^2} \, H_T=0 \;,
\ee

\be  \label{31}
- \dot{\Phi}_H +\left( H +  \frac{\dot{F}}{2F} \right) \Phi_A = 
\frac{1}{2} \left(  \frac{  \Delta \dot{F} }{F} -H \frac{ \Delta F}{F} +  
\frac{ \omega}{F} \,  \dot{\phi} \, \Delta \phi \right)  \; ,
\ee

\begin{eqnarray}  \label{32}
& & \left( \frac{k}{a} \right)^2 \Phi_H +\frac{1}{2}
\left( \frac{ \omega}{F } \dot{\phi}^2  + \frac{3}{2} \frac{\dot{F}^2}{F^2}  \right) \Phi_A =
\frac{1}{2} \left\{ \frac{3}{2} \frac{ \dot{F} \Delta \dot{F} }{F^2} + \left( 3\dot{H} -  
\frac{k^2}{a^2} -\frac{3H}{2} \frac{ \dot{F}}{F}  \right)  \frac{ \Delta F}{F} 
\right. \nonumber \\
&& \nonumber \\
& & \left. +\frac{\omega}{F} \dot{\phi} \Delta \dot{\phi} + \frac{1}{2F} \left[ 
\dot{\phi}^2 
\frac{d\omega}{d\phi} -\frac{ \partial f}{\partial \phi} +2\frac{dV}{d\phi} +6\omega \dot{\phi} 
\left( H +  \frac{ \dot{F} }{2F} \right) \right] \Delta \phi \right\} 
\; ,
\end{eqnarray}

\be   \label{33}
\Phi_A + \Phi_H =  - \frac{\Delta F }{F} \; ,
\ee

\begin{eqnarray}  
& & \ddot{\Phi}_H +H \dot{\Phi}_H + \left( H + \frac{ \dot{F}}{2F} \right) 
\left( 2\dot{\Phi}_H -\dot{\Phi}_A \right) 
+\frac{ 1 }{2F}  \left( f-2V -RF \right)  \Phi_A \nonumber \\ 
&& \nonumber \\
& & =  - \frac{1}{2} \left[ 
\frac{  \Delta \ddot{F}}{F} + 2H \, \frac{\Delta \dot{F}}{F} 
+ \left( P-\rho \right) \frac{ \Delta F}{2F} + \frac{ \omega}{F} \, \dot{\phi} \, \Delta 
\dot{\phi} 
+ \frac{1}{2F} \left( \dot{\phi}^2 \, \frac{d\omega}{ d\phi}  +\frac{\partial 
f}{\partial \phi 
}  -2 \, \frac{dV}{d\phi } \right) \Delta \phi  \right]  \; ,\nonumber \\
&& \nonumber \\
&& \label{34}
\end{eqnarray}
where $\Delta \dot{F} \equiv d( \Delta F) / dt$, {\em etc.}, 
\be \label{35}
\Delta R=6 \left[ \ddot{\Phi}_H +4H\dot{\Phi}_H +\frac{2}{3} \frac{k^2}{a^2} \Phi_H -H\dot{\Phi}_A 
-\left( 2\dot{H}+4H^2 -\frac{k^2}{3a^2} \right) \Phi_A \right] \;, 
\ee
and the effective energy density and pressure of the scalar are given by
\be  \label{36}
\rho =\frac{1}{F}  \left[  \frac{\omega \, \dot{\phi}^2}{2} +\frac{1}{2} 
\left( RF -f +2V \right) -3H \dot{F} +  \nabla^c F_c \right] \; ,
\ee

\be  \label{37}
P=\frac{1}{F}  \left[  \frac{\omega \, \dot{\phi}^2}{2} +\frac{1}{2} 
\left( f -RF-2V \right) +\ddot{F} +2H \dot{F} -\frac{2}{3} \, \nabla^c F_c \right] \; .
\ee

Here $ F_c \equiv h_c^d \, \nabla_d F$ is the spatial projection of the gradient of $F$. In  
the de 
Sitter background (\ref{desitter}) the 
gauge-invariant variables reduce, to first order, to 
\be \label{38}
\Delta \phi =\delta \phi \;, \;\;\;\; \Delta R=\delta R\; , \;\;\; \Delta F=\delta F \;, \;\;\;\; 
\Delta f =\delta f \;,
\ee
and the equations they obey reduce, to first order, to
\be  \label{42}
\Delta \ddot{\phi} + 3H_0  \Delta \dot{\phi} 
+ \left[ \frac{k^2}{a^2}-\,  \frac{1}{2\omega_0} \left( f_0''   - 2 V_0'' \right) \right] 
\Delta \phi  =
 \frac{ f_{\phi R}}{2 \omega_0 } \,  \Delta R  \; ,
\ee

\be  \label{43}
\Delta \ddot{F} +3H_0 \, \Delta \dot{F} +\left( \frac{k^2}{a^2} - 4H_0^2 \right) \Delta F 
+\frac{F_0}{3} \, \Delta R =0 \;,
\ee

\be \label{44}
\ddot{H}_T +3H_0  \, \dot{H}_T +\frac{k^2}{a^2} \, H_T=0 \;,
\ee

\be \label{45}
-\dot{\Phi}_H+H_0 \Phi_A =\frac{1}{2} \left( \frac{\Delta \dot{F}}{F_0} -H_0 \, \frac{ \Delta 
F}{F_0} 
\right) \;,
\ee

\be  \label{46}
\Phi_H  = -  \frac{1}{2} \, \frac{ \Delta F}{F_0}   \; ,
\ee

\be   \label{47}
\Phi_A + \Phi_H =  - \frac{\Delta F }{F_0} \; ,
\ee

\be  \label{48}
\ddot{\Phi}_H + 3H_0 \dot{\Phi}_H  - H_0  \dot{\Phi}_A -3H_0^2 \Phi_A  
 =  - \frac{1}{2}\frac{  \Delta \ddot{F}}{F_0} - H_0 \, \frac{\Delta \dot{F}}{F_0} 
+ \frac{3H_0^2 }{2} \, \frac{\Delta F}{ F_0 }   \; ,
\ee
with 
\be \label{49}
\Delta R=6 \left[ \ddot{\Phi}_H + 4H_0 \dot{\Phi}_H + \frac{2}{3} \frac{k^2}{a^2} \, \Phi_H 
-H_0 \dot{\Phi}_A + \left( \frac{k^2}{3a^2} -4H_0^2 \right) \Phi_A \right] \;,
\ee
and where
\be
 f_{\phi R} \equiv \left.  \frac{ \partial^2 f}{ \partial \phi \partial R} 
\right|_{ \left( \phi_0, R_0 \right)} \;, \;\;\;\;\;\;
 f_{R R} \equiv \left.  \frac{ \partial^2 f}{ \partial R^2} \right|_{ \left( \phi_0, R_0 
\right)} \;, \;\;\;\;\;\; f_0'' \equiv \left. \frac{\partial^2 f}{\partial \phi^2} 
\right|_{\left( \phi_0, R_0 \right)} \;.
\ee

%%%%%%%%%%%%
The comparison of eqs.~(\ref{46}) and (\ref{47}) yields
\be \label{50}
\Phi_H=\Phi_A =-\, \frac{\Delta F}{2F_0} 
\ee
which, substituted in eq.~(\ref{49}), leads to
\be \label{52}
\Delta R=6 \left[ \ddot{\Phi}_H + 3H_0 \dot{\Phi}_H + \left( \frac{k^2}{a^2}  -4H_0^2 \right) \Phi_H 
 \right] \;.
\ee
Eqs.~(\ref{42}) and (\ref{44}) do not change form, while the remaining equations reduce to identities.
The decoupling of scalar, vector and tensor modes is not apparent in the formalism used, 
with the exception of tensor modes described by the perturbation $H_T$. This quantity obeys 
eq.~(\ref{44}), which is decoupled from the other modes. We do not consider vector modes 
described by the quantity $B$ since, as proven in Ref.~\cite{HwangCQG}, vorticity modes 
cannot be generated in generalized gravity when matter contributions are absent (i.e., when the 
scalar field or nonlinear corrections to $R$ dominate). The vector mode $B$ effectively disappears
from the gauge-invariant variables $ \Delta \phi$ and $\Phi_A=\Phi_H=-\Delta F/ (2F_0)$ defined 
by mixing scalar modes and $B$. More naively, other authors refer to these facts by saying
 that the vector perturbations can be gauged away. For this reason in the following we consider 
explicitly only  scalar 
and tensor perturbations.

%%%%%%%%%%%%%%%%%%%%%
\subsection{Stability with respect to tensor perturbations}

The evolution of tensor perturbations is regulated by eq.~(\ref{44}), 
where $a(t)=a_0 \, \mbox{e}^{H_0\, 
t} $. By introducing the auxiliary variable
\be \label{54}
u \equiv a\, H_T
\ee
and using conformal time $\eta$ defined by $dt=a \, d\eta$ and the standard relation 
\be  \label{57}
\mbox{e}^{ H_0 \, t} =-\, \frac{1}{ a_0 \, H_0 \, \eta} 
\ee
valid in de Sitter space, eq.~(\ref{44}) is 
reduced to the formal Schr\"{o}dinger equation
\be \label{58}
\frac{d^2 u}{d\eta^2} +\left( k^2 - \frac{ 2}{\eta^2}  \right) u ( \eta )=0 \;.
\ee

Let us consider the expanding ($H_0 >0$) de Sitter spaces (\ref{desitter}). We are interested in the 
late time evolution of perturbations, corresponding to $t \rightarrow +\infty$ and 
$ \eta 
\rightarrow 0^{-}$. In this regime, the general solution of the asymptotic equation
\be \label{60}
\frac{d^2 u}{d\eta^2} -  \frac{ 2}{\eta^2} \, u ( \eta )=0 
\ee
is
\be \label{61}
u( \eta )= \frac{C_1}{\eta}+ C_2 \, \eta^2 
\ee
for $\eta \neq 0$, where $C_{1,2}$ are integration constants. Then 
\be\label{62}
H_T=-H_0 \left( C_1 +C_2 \, \eta^3 \right) \;,
\ee
and the gauge-invariant tensor perturbation doesn't grow when $\eta \rightarrow 0^{-}$. 
Hence, expanding de Sitter spaces are always stable 
with respect to tensor perturbations.

Let us consider also the contracting ($H_0 < 0$) de Sitter spaces (\ref{desitter}). In this case 
$ t 
\rightarrow +\infty $ corresponds to $\eta \rightarrow +\infty $ and the asymptotic solutions of 
eq.~(\ref{58}) are free waves $\mbox{e}^{\pm ik \eta} $, hence the amplitude of the tensor 
perturbations
\be \label{63}
H_T=\frac{ \mbox{e}^{\pm ik \eta} }{a_0} \,  \mbox{e}^{ \left| H_0 \right| \, t}
\ee
diverges when $t \rightarrow +\infty $. As a conclusion, contracting de Sitter 
spaces are always 
unstable with respect to tensor perturbations.

%%%%%%%%%%%%%%%%%%%%%%%%%%
\subsection{Stability with respect to scalar perturbations}

At a first glance it might seem that we are left with only one equation (\ref{42}) to 
determine the 
perturbations $\Delta \phi$ 
and $\Phi_H=\Phi_A=-\Delta F/ (2F_0) $, but this is not the case since one can Taylor-expand the 
coupling function $ f \left( \phi, R \right)$ obtaining 
\be \label{64}
\frac{ \Delta F}{F_0} = \frac{ f_{\phi R}}{F_0} \, \Delta \phi +
\frac{ f_{R R}}{F_0} \, \Delta R \;.
\ee
Eqs.~(\ref{50}) and (\ref{64}) then yield 
\be \label{65}
\Delta R= \frac{-2F_0}{f_{RR}} \, \Phi_H  -\, \frac{ f_{\phi R} }{ f_{RR}} \, \Delta \phi \;,
\ee
while eq.~(\ref{42}) becomes 
\be  \label{66}
\Delta \ddot{\phi} + 3H_0  \Delta \dot{\phi} 
+ \left[ \frac{k^2}{a^2}-\,  \frac{1}{2\omega_0} \left( f_0''   - 2 V_0'' 
-\, \frac{f_{\phi R}^2}{f_{RR}}  
\right) \right]  \Delta \phi  + \frac{F_0}{\omega_0} \, \frac{ f_{\phi R}}{ f_{RR} }  
\, \Phi_H =0 \; .
\ee
The comparison of eqs.~(\ref{65}) and (\ref{49}) yields
\be  \label{67}
\ddot{\Phi}_H + 3H_0 \dot{\Phi}_H  +\left( \frac{k^2}{a^2} -4H_0^2 +\frac{F_0}{3f_{RR}} \right) 
\Phi_H 
+ \frac{ f_{\phi R}}{ 6 f_{RR} } \, \Delta \phi =0   \; .
\ee
In the rest of this section we consider the case in which $f_{RR}\neq 0$. This restriction leaves 
out 
linear theories of gravity, including general relativity, which are discussed in the next section.

We have now the system (\ref{66}) and (\ref{67}) for $ \Delta \phi $ and $ \Phi_H 
$, which 
can be simplified by switching to the variables $v$ and $w$ defined by 
\be \label{68}
\Phi_H \equiv \frac{v}{a} \;, \;\;\;\;\;\;\;\;
\Delta \phi  \equiv \frac{w}{a} \;,
\ee
and by using conformal time $\eta $ instead of $t$. In terms of these new variables it is
\be  \label{71}
\frac{ d^2 v}{d\eta^2 }  + \left( k^2 + \frac{\alpha}{\eta^2} \right) v
+ \frac{\beta }{\eta^2} \, w =0 \; ,
\ee
\be  \label{72}
\frac{ d^2 w}{ d\eta^2 }  + \left( k^2 + \frac{\gamma}{\eta^2} \right) w
+ \frac{\delta  }{\eta^2} \, v =0 \; ,
\ee
where 
\begin{eqnarray}
\alpha & = & \frac{F_0}{ 3 f_{RR} H_0^2} -6 \;, \\
&& \nonumber \\
\beta & = &  \frac{ f_{\phi R} }{ 6 f_{RR} H_0^2} \;,\\
&& \nonumber \\
\gamma & = &  \frac{1}{2\omega_0} \left( 2V_0'' -f_0'' + \frac{ f_{\phi R}^2 }{ f_{RR}}  \right)  -2 
\;,\\
&& \nonumber \\
\delta & = &  \frac{F_0}{ \omega_0 H_0^2} \, \frac{ f_{\phi R} }{ f_{RR}}  \;.
\end{eqnarray}

The system (\ref{71}) and (\ref{72}) can be linearized around the conformal time $\eta_0$ at 
which the perturbations originate and then rewritten as the first order system
\begin{eqnarray}
 v' & \equiv & x  \;,\\
&& \nonumber \\
 w' & \equiv & y \;,\\
&& \nonumber \\ 
x' & = &   - \left( k^2 + \frac{\alpha}{\eta_0^2}  \right) v
- \frac{ \beta}{\eta_0^2}  \, w  \; ,\\
&& \nonumber \\
y' & = &    -  \left( k^2 + \frac{\gamma}{\eta_0^2} \right) w -\, \frac{\delta}{\eta_0^2}   \, v  \;.
\end{eqnarray}
This can be written in compact form as 
\be   \label{birba}
\left( 
\begin{array}{c}
 v'  \\
 w' \\
 x' \\
 y' 
\end{array} \right) = \hat{M}  
\left( 
\begin{array}{c}
v \\
w\\
x\\
y
\end{array} \right) \;,
\ee
where the matrix $ \hat{M} $ is 
\be
\hat{M} =\left(
\begin{array}{cccc}
0 & 0 & 1 & 0 \\
&&&\\
0 & 0 & 0 & 1 \\
&&&\\
- \left( k^2 +\frac{\alpha}{\eta_0^2} \right)  &
- \frac{  \beta }{ \eta_0^2 }   & 0  & 0  \\
&&&\\
- \,\frac{ \delta }{ \eta_0^2  }  &
-\left( k^2+ \frac{\gamma}{\eta_0^2} \right)   & 0  & 0  
\end{array} \right) \;.
\ee
A stability analysis for this system (see Appendix A) yields the result that de Sitter space is 
stable if any one of the following conditions is satisfied:

\begin{itemize} 

\item if $ b_1 \geq 0$ and $c_1 \geq 0$, it must be $ k\geq k_1 $ for 
stability

\item if $ b_1 > 0$ and $c_1 < 0$, it must be $ k\geq \mbox{max}\left\{ k_1, k_3 \right\} $

\item if $ b_1 = 0$ and $c_1 \geq 0$, it must be $ k\geq  k_1 $

\item if $ b_1 = 0$ and $c_1 < 0$, it must be $ k\geq \mbox{max}\left\{ k_1, k_4\right\} $

\item if $ -2 \leq  b_1 < 0$ and $c_1 > b_1^2/4 $, it must be $ k\geq  k_1 $

\item if $ -2 \leq b_1 < 0$ and $ 0 < c_1 < b_1^2/4  $, it must be 
$  k_1 \leq k \leq k_5 $ (this inequality can be satisfied only if $k_5 > k_1  $) or
 $ k \geq  \mbox{max}\left\{ k_1, k_6 \right\} $

\item if $ -2 \leq b_1 <  0$ and $c_1 < 0$, it must be $ k \geq \mbox{max}\left\{ k_1,  
k_7 \right\} $

\item if $  b_1 <  -2 $ and $c_1 < 0$, it must be $ k\geq \mbox{max}\left\{ k_1, k_2, k_7  
\right\} $

\item if $ b_1 < -2 $ and $ 0< c_1 \leq b_1^2 /4  $, it must be $ \mbox{max} 
\left\{ k_1, k_2 \right\} \leq k \leq k_7 $ (when $k_7 > k_1, k_2$) or  $ k \geq 
\mbox{max}\left\{ k_1, k_2, k_6  
\right\} $

\item if $ b_1 <  -2 $ and $c_1 > b_1^2/4 $, it must be $ k \geq \mbox{max}\left\{ k_1, 
k_2 \right\} $.

\end{itemize} 

The parameters $b_1, c_1$ and the critical wave vectors $k_i$ are defined in terms of the values of 
the coupling functions and parameters by 
\begin{eqnarray}
b_1 & = & \frac{1}{ \eta_0^2} \left[ \frac{1}{2\omega_0} \left( 2V_0''-f_0''
+\frac{ f_{\phi R}^2}{f_{RR}} \right) + \frac{ F_0 }{3 f_{RR} H_0^2} -8  \right] -2
\;, \\
&& \nonumber \\
&& \nonumber \\
c_1 & = & \frac{1}{ 4 \eta_0^4} \left[ 
\frac{1}{2\omega_0} \left( 2V_0''-f_0''+\frac{ f_{\phi R}^2}{f_{RR}} \right) -8 
+\frac{ F_0}{3f_{RR} H_0^2} \right]^2  \nonumber \\
&& \nonumber \\
&&   - \frac{1}{\eta_0^2}  \left[ \frac{ F_0}{3f_{RR}H_0^2} 
+ \frac{1}{2\omega_0} \left( 2V_0'' -f_0'' + \frac{ f_{\phi R}^2 }{f_{RR}} \right) -8 \right] +\, 
\frac{ F_0 f_{\phi R}^2 }{6 \omega_0 f_{RR}^2 H_0^4 \eta_0^4 }  \;,
\end{eqnarray}

\be 
k_1 = \left\{ 
\frac{1}{2\eta_0^2} \left[ \frac{1}{2\omega_0} \left( f_0''-2V_0'' -\,\frac{ f_{\phi R}^2}{ f_{RR}} 
+ \frac{ F_0 f_{\phi R}^2}{ 3H_0^4 f_{RR}^2 } \right) +8 -\,\frac{F_0}{3f_{RR}H_0^2} 
\right] \right\}^{1/2} 
\;,
\ee

\be
k_2 = \left\{ 
\frac{1}{2\eta_0^2} \left[ \frac{1}{2\omega_0} \left( f_0''-2V_0'' -\,\frac{ f_{\phi R}^2}{ f_{RR}} 
\right) + 8 - \frac{ F_0 }{ 3H_0^2 f_{RR} }  \right] \right\}^{1/2} 
\;,
\ee

\be
k_3= \sqrt{ \frac{ -b_1 + \sqrt{ b_1^2-4c_1}}{2} } \;,
\ee

\be
k_4 = \left| c_1 \right|^{1/4} \;,
\ee

\be
k_{5,6} = \sqrt{ \frac{  \left| b_1 \right| \pm \sqrt{ b_1^2-4c_1}  }{2} } \;,
\ee

\be
k_7 =
 \sqrt{ \frac{  \left| b_1 \right| + \sqrt{ b_1^2 + 4 \left| c_1 \right| }  }{2} } \;.
\ee

The effective Jeans wavelengths $\lambda_{i} = 2\pi/k_{i} $ are determined by the values of 
$f(\phi, R ) , V,$ and $ \omega $ at the point $\left( \phi_0, R_0 \right)$ and by the value of the 
initial 
conformal time $\eta_0$. The conditions for stability, when expressed in 
terms of $ f,V$  and $\omega_0, \eta_0 $ are not particularly illuminating -- they become more 
meaningful when  specific forms of these functions are adopted. Although the 
classification of the stability regions seems involved, once the generalized  gravity theory is 
specified, the values of the parameters $b_1$ and $c_1 $ are completely fixed and only one of the 
cases contemplated in the list above applies. Therefore this list contains complete conditions 
to check at a glance whether  de Sitter space is stable against inhomogeneous perturbations in 
general theories of gravity.

As an example consider the theory described by
\be
f=R+\epsilon R^2  \;, \; \;\;\; \phi =1 \;, \;\;\;\;\omega=1\;,\;\;\;\; V=\Lambda > 0 \;,
\ee
where $\epsilon$ is a small positive constant (general relativity is recovered in the limit $
\epsilon \rightarrow 0$). One has
\be
f_0=12H_0^2 \left( 1+12\epsilon H_0^2 \right) \;, \;\;\; F_0=1+24\epsilon H_0^2 \;, 
\;\;\;\; V_0=\Lambda \;, 
\ee
\be
 V_0'=f_0'=f_0''=f_{\phi R}=f_{\phi\phi}=0 \;, \;\;\;\; f_{RR}=2\epsilon
\ee
and $H_0=\sqrt{ \Lambda/3} $ as in general relativity. Since $\alpha=
\left( 2\epsilon \Lambda \right)^{-1} \, -2$ and $\beta=\gamma=\delta=0$, 
eqs.~(\ref{71}) and (\ref{72}) reduce, in the limit $\epsilon \rightarrow 0$, to
\be  
v''+\left( k^2 +\frac{1}{2\epsilon \Lambda \eta_0^2} \right) v=0 \;, \label{dddelta}
\ee
\be  \label{dddeltadelta}
w''+k^2 w=0 \;,
\ee
which constitute decoupled equations for the variables $v$ and $w$ with positive 
(angular) frequency squared and hence describe oscillatory perturbations associated with stability
of de Sitter space. The small correction $\epsilon R^2$ to the Einstein-Hilbert Lagrangian 
does not destabilize de Sitter space. Naively, this correction ``reinforces'' the effects of $R$.
If, however, $\epsilon$ were allowed to be negative, the effect of $\epsilon R^2$ 
would be the opposite and it would tend to destabilize de Sitter space. 
This is evident in eq.~(\ref{dddelta}) when $\epsilon <0$ and the 
term containing $\epsilon$ dominates for $\epsilon \rightarrow 0$, making 
the effective frequency squared of $v$ negative and introducing  exponential 
solutions associated with instability.

%%%%%%%%%%%%%%%%%%%%%%%%%%%%%%%%%%%%%%%%%%%%%%%%%%%%%%%%%%%%%%%%%%%%%%%%%%%%%%%%%%%%

\section{Scalar-tensor theories}
\setcounter{equation}{0}

We now restrict the stability analysis to scalar-tensor theories, for 
which the coupling $f\left( \phi, R \right)$ is linear in $R$ and $f_{RR}=0$ -- this case was 
not contemplated in Sec.~3. Having already established 
that contracting de Sitter spaces are always unstable with respect to tensor perturbations, we only 
need to consider an expanding de Sitter background (\ref{desitter}) with $H_0>0$. It is 
straightforward to derive the first order evolution equations for the perturbations. Eqs.~(\ref{50}), 
(\ref{64}) and (\ref{65}) yield 
\be \label{87}
\Phi_H= -\, \frac{1}{2} \, \frac{f_{\phi R}}{F_0} \, \Delta \phi 
\ee
while eqs.~(\ref{52}) and (\ref{87}) lead to 
\be \label{88}
\Delta R= -\, \frac{3 f_{\phi R} }{ F_0 } \left[  \Delta \ddot{\phi} +3H_0 \, \Delta 
\dot{\phi} + \left( \frac{k^2}{a^2} -4H_0^2 \right) \Delta \phi \right] \;.
\ee
In conjunction with eq.~(\ref{42}) this yields
\be \label{89}
\Delta \ddot{\phi} + 3H_0 \, \Delta \dot{\phi} 
+ \left[ \frac{k^2}{a^2}-\,  \frac{  \left( \frac{f_0''}{2}   -  V_0'' +\frac{6 f_{\phi R}^2}{F_0} \, 
H_0^2 \right)}{\omega_0 \, \left( 1+\frac{ 3f_{\phi R}^2}{2\omega_0 F_0} \right)}   \right] 
\Delta \phi  = 0  
\ee
if $ 1+3f_{\phi R}^2/( 2 \omega_0 F_0) \neq 0$. In the case in which 
$ 1+3f_{\phi R}^2/( 2 \omega_0 F_0) = 0 $, instead, eq.~(\ref{42}) yields either the trivial 
solution $\Delta \phi=0$ or $ f_0''-2V_0''=8\omega_0 \, H_0^2$ 
\cite{lastfootnote}.

We look for solutions of the asymptotic form of eq.~(\ref{89}) at late times satisfying the ansatz
\be \label{90}
\Delta \phi =\epsilon \, \mbox{e}^{s t} \;,
\ee
where $\epsilon $ and  $s$ are constants, with $s$  satisfying the algebraic equation
\be \label{91}
s^2 +3H_0 \, s + c =0 \;,
\ee
with
\be \label{91bis}
c = -\, \frac{  \left( \frac{f_0''}{2}   -  V_0'' +\frac{6 f_{\phi R}^2}{F_0} \, 
H_0^2 \right)}{\omega_0 \, \left( 1+\frac{ 3f_{\phi R}^2}{2\omega_0 F_0} \right)} \;.
\ee
The roots 
\be 
s_{\pm} =\frac{ -3H_0 \pm \sqrt{ 9H_0^2-4c}}{2} 
\ee
of eq.~(\ref{91bis}) are such that $Re\left( s_{-} \right) <0$, while the sign of $Re\left( s_{+} \right)$ 
depends on the sign of~$c$. If $c \geq 0$ then $Re\left( s_{+} \right) \leq 0 $ and there is 
stability. If 
instead  $c < 0$ then $ Re\left( s_{+} \right) > 0 $ and the de Sitter space (\ref{desitter}) is unstable. 
The condition for the stability of de Sitter space in a scalar-tensor theory described by the action 
(\ref{I9}) is then
\be\label{93}
\frac{  \left( \frac{f_0''}{2}   -  V_0'' +\frac{6 f_{\phi R}^2}{F_0} \, 
H_0^2 \right)}{\omega_0 \, \left( 1+\frac{ 3f_{\phi R}^2}{2\omega_0 F_0} \right)} \leq 0 \;.
\ee

\vskip1truecm
\noindent $ \bullet $  {\em Non-minimally coupled scalar field}\\

As a particular case of scalar-tensor gravity we  consider the theory of a non-minimally 
coupled 
scalar field given by the choice (\ref{NMC}) of the coupling functions, yielding
\be  \label{98}
F_0=1-\xi \phi_0^2 \;, \;\;\;\;\; f_0'=-2\xi R_0 \phi_0=2V_0' \;, \;\;\;\;\;
f_{\phi R}=-2\xi \phi_0 =\frac{V_0'}{6H_0^2} \;, \;\;\;\;\; f_0''=\frac{2V_0'}{\phi_0} \;,
\ee
where eq. (\ref{15}) has been used. If the effective gravitational coupling of the theory  
\be  \label{effectiveG}
G_{eff} \equiv \frac{G}{1-8\pi G \xi \phi^2} 
\ee
is positive (which happens for any negative value of $\xi$ or, if $\xi>0$, for $\left| \phi 
\right| < \left( 8\pi G \xi \right)^{-1/2} $), then the denominator on the left hand side of 
eq.~(\ref{93}) is also positive and the stability condition of de Sitter space reduces to 
\be \label{99}
V_0'' -\frac{f_0''}{2} -\frac{6 f_{\phi R}^2 H_0^2 }{F_0} \geq 0 \;.
\ee
Upon use of eq.~(\ref{98}), this condition is written as 
\be \label{104}
V_0'' \geq f( x) \,\, \frac{V_0'}{\phi_0} 
\;,
\ee
where 
\be \label{105}
x= \xi \phi_0^2 \;, \;\;\;\;\; f(x)=\frac{ 1-3x}{1-x} \;.
\ee
Eq.~(\ref{104}) coincides with the stability condition  
found in Refs.~\cite{FaraoniPLA,FaraoniIJTP} when $\phi_0\neq 0$ \cite{footnote3}. 
If the effective coupling (\ref{effectiveG}) is instead negative, the stability condition is 
given by (\ref{104}) with the direction reversed.

In the 
case  $\phi_0=0$ one has $f_0=R_0=12H_0^2 , F_0=1,f_0'=f_{\phi R}=0, f_0''=-2\xi R_0$, and the 
condition for the stability of de Sitter space becomes
\be \label{112}
V_0''+12\xi H_0^2 \geq 0 \;.
\ee
Using eq.~(\ref{14}), this assumes the form
\be
V_0''+4\xi V_0 \geq 0 \;,
\ee
the stability condition found in Ref.~\cite{FaraoniPLA,FaraoniIJTP} for $\phi_0=0$. Eq. (\ref{93})  
generalizes to arbitrary scalar-tensor theories the stability conditions already known for 
non-minimally coupled scalar field  theory, which are recovered as a special case.

\vskip1truecm
\noindent $ \bullet $ {\em General relativity}\\

In Einstein gravity with a minimally coupled scalar, 
\be
\omega = F_0=1 \; ,\;\;\;\;\;\;  f_0=12H_0^2 \; , \;\;\;\;\;\; f_0'=f_0''=f_{\phi 
R}=f_{RR}=0 \;,
\ee
and the de Sitter space obtained if $ H_0=\sqrt{V_0/3} $,~~$ V_0'=0$ is stable if $ 
V_0''  \geq 0$, in particular if the potential has a minimum at $\phi_0$. Hence in this case the 
concavity of the scalar field potential is the stabilizing factor, while its convexity would 
instead cause instability.
 
If the scalar is absent the de Sitter space obtained thanks to a positive cosmological constant is 
automatically guaranteed to be stable with respect to inhomogeneous perturbations. It is well known 
that this space is also stable with respect to large anisotropic perturbations, with the exception 
of highly positively curved Bianchi IX models, as 
described by the cosmic no-hair theorems \cite{cosmicnohair}.

\vskip1truecm
\noindent $ \bullet $ {\em Phantom field}\\

The superstring-inspired theory of a phantom field with negative kinetic energy corresponds to 
$ f(\phi, R)=R$ and $\omega=-1$.  The conditions for the existence of de Sitter solutions are
\be
H_0^2= \frac{ V_0}{3} \;, \;\;\;\;\;\; \;\;\; V_0'=0 \; ,
\ee
while the condition for stability reduces to $ V_0'' \leq 0$. Thus, de Sitter fixed points 
$ \left( H_0, \phi_0 \right) $ are attractors if $V_0'' $ has a maximum at $\phi_0$. Due to 
the negative sign of its kinetic energy the phantom field $\phi$ ``falls up'' and settles in 
the maximum of the potential. de Sitter attractors have been found in superaccelerating models 
of dark energy, thus avoiding evolution of the universe in a Big Rip singularity in the future 
\cite{Singhetal,SamiToporensky,HaoLi}.

%%%%%%%%%%%%%%%%%%%%%%%%%%%%%%%%%%%%%%%%%%%%%%%%%%%%%%%%%%%%%%%%%%%%%%%%%%%%%%%%%%%%

\section{Minkowski space and its stability}
\setcounter{equation}{0}

Generalized gravity often admits Minkowski space solutions. In general, flat space solutions in 
generalized gravity are physically  non-trivial and correspond 
to a balance between gravity and the  scalar field $\phi$ formally acting as a material source.  
In non-linear gravity theories where this balance cannot be achieved, 
a Minkowski solution may not exist. This is the case, for  example, of the theory described by 
eq.~(\ref{questa}).

It has been suggested that the present universe 
could have originated from Minkowski space 
\cite{GunzigetalPRD,SaaetalIJTP} or from a static Einstein space 
\cite{EllisMaartens02,Barrowetal03}. To pursue 
this idea it is necessary to ascertain the stability of Minkowski or Einstein space. The stability 
of Einstein spaces is  a long-standing issue \cite{Eddington30,Harrison67,Gibbons} 
and it has recently been revisited  in 
general relativity by  considering inhomogeneous and anisotropic perturbations 
\cite{Barrowetal03}. Here we consider the stability of Minkowski space in generalized gravity with 
respect to inhomogeneous perturbations.

The conditions for the existence of Minkowski solutions (with $H_0=0$) of  
the field equations of generalized gravity are
\be   \label{10051}
f_0-2V_0=0 \;,
\ee
\be  
f_0'-2V_0'=0 \;.
\ee
A positive value of $f_0$, which describes a realistic situation, can be balanced 
in eq.~(\ref{10051}) by a negative cosmological constant $\Lambda$, which is familiar to high energy 
physicists working with anti-de Sitter space. 

In order to study the stability of Minkowski space one needs the linearized equations for the 
gauge-invariant variables
\be  \label{1000}
\Delta \ddot{\phi} + \left[ k^2-\,  \frac{1}{2\omega_0} \left( f_0''   - 2 V_0'' \right) 
\right]  \Delta \phi  =
 \frac{ f_{\phi R}}{2 \omega_0 } \, \Delta R  \; ,
\ee

\be \label{1001}
\Delta \ddot{F}   + k^2  \Delta F  +\frac{F_0}{3} \, \Delta R =0 \;,
\ee

\be \label{1002}
\ddot{H}_T  + k^2 \, H_T=0 \;,
\ee

\be \label{1003}
\dot{\Phi}_H  = - \, \frac{1}{2} \, \frac{\Delta \dot{F}}{F_0}  \;,
\ee

\be   \label{1004}
\Phi_A + \Phi_H =  - \frac{\Delta F }{F_0} \; .
\ee

Eq.~(\ref{1003}) yields again  $ \Phi_H = -\Delta F / (2F_0) = \Phi_A $, while eq.~(\ref{49}) reduces 
to 
\be
\Delta R = 6 \left( \ddot{\Phi}_H +k^2 \, \Phi_H \right) \;.
\ee

Again, we consider scalar and tensor modes and drop the vector modes which cannot 
be generated in the absence of matter \cite{HwangCQG}.
Inspection of eq.~(\ref{1002}) allows one to conclude at once that Minkowski space is always stable 
against tensor perturbations, which decouple from the other modes. 
To assess stability with respect to scalar perturbations, one considers 
the analogue of eqs.~(\ref{66}) and (\ref{67}), which are 
\be  \label{1005}
\frac{ d^2 v}{d t^2 }  + \left( k^2 + \frac{F_0}{ 3f_{RR} } \right)  v
+ \frac{ f_{\phi R} }{ 6f_{RR} }  \, w =0 \; ,
\ee

\be  \label{1006}
\frac{ d^2 w}{ d t^2 }  +\left[ k^2 + \frac{ 1}{2\omega_0} \left( 2V_0'' -f_0'' +\frac{f_{\phi 
R}^2}{f_{RR} } \right) \right] w + \frac{ F_0  }{\omega_0 }\, \frac{ f_{\phi R }}{ f_{RR} }   \, v =0 
\ee
in the case $f_{RR} \neq 0$, where the variables $v$ and $w$  introduced in eq.~(\ref{68}) now 
coincide with $\Phi_H=\Phi_A $ and $ \Delta \phi$, respectively.

The system (\ref{1005}) and (\ref{1006}) can be reformulated as the first order system
\begin{eqnarray}
 v' & \equiv & x  \;,\\
&& \nonumber \\
 w' & \equiv & y \;,\\
&& \nonumber \\ 
x' & = &   - \left( k^2 +\frac{F_0}{ 3f_{RR} }\right)  v
- \frac{ f_{\phi R} }{ 6f_{RR} }  \, w  \; ,\\
&& \nonumber \\
y' & = &    -  \frac{ F_0 \, f_{\phi R } }{\omega_0 \,f_{RR} }   \, v  
+D  w \;, 
\end{eqnarray}
where
\be
D = - \left[ k^2 +  \frac{ 1}{2\omega_0} \left( 2V_0'' -f_0'' +\frac{f_{\phi R}^2}{f_{RR} 
}  \right) \right] \;.
\ee
In compact form,  
\be   \label{1007}
\left( 
\begin{array}{c}
 v'  \\
 w' \\
 x' \\
 y' 
\end{array} \right) = \hat{N} 
\left( 
\begin{array}{c}
v \\
w\\
x\\
y
\end{array} \right) \;,
\ee
where  
\be
\hat{N} = \left(
\begin{array}{cccc}
0 & 0 & 1 & 0 \\
&&&\\
0 & 0 & 0 & 1 \\
&&&\\
- \left( k^2 + \frac{F_0 \, f_{\phi R} }{ 3f_{RR} } \right)  &
- \frac{ f_{\phi R}}{6f_{RR} }   & 0  & 0  \\
&&&\\
- \frac{F_0 \, f_{\phi R} }{ \omega_0 \, f_{RR} }  &
D   & 0  & 0  
\end{array} \right) \;.
\ee
A stability analysis for this linear system, presented in Appendix B, leads to 
the result that Minkowski space is stable if $ 4c_2 \leq b_2^2 $ and one of the following conditions 
is satisfied:

\begin{itemize} 

\item $ b_2 = 0$, $c_2 \leq 0$, and $k\geq \mbox{max} \left\{ k_8, k_{10} \right\} $

\item $ b_2 < 0$,  $ c_2 < 0$, and $ k\geq \mbox{max}\left\{ k_8,  k_{13} \right\} $

\item $ b_2 > 0$ and  $c_2 \geq 0$

\item $ b_2 > 0$, $c_2 < 0$, and  $ k\geq k_9  $,

\end{itemize} 

where 
\begin{eqnarray}
b_2 & = & \frac{1}{2\omega_0} \left( 2V_0'' -f_0'' +\frac{ f_{\phi R}}{f_{RR}} \right)
+ \frac{ F_0 f_{\phi R} }{ 3f_{RR} } \;, \\
&& \nonumber \\
c_2 & = &  \frac{F_0 f_{\phi R}}{6\omega_0 f_{RR}}  \left( 
2V_0'' -f_0''  \right)  \;, \\
&& \nonumber \\
k_8 & = & \sqrt{ \frac{ |b_2|}{2}} \; , \\
& & \nonumber \\ 
k_9 & = &  \sqrt{ \frac{ -b_2 + \sqrt{ b_2^2-4c_2}}{2} } \;, \\
&& \nonumber \\
k_{10} & = & \left| c_2 \right|^{1/4} \; , \\
&& \nonumber \\
k_{13} & = &  \sqrt{ \frac{  \left| b_2 \right| + \sqrt{ b_2^2 + 4 \left| c_2 \right| }  }{2} } \;.
\end{eqnarray}

Consider as examples general relativity and the theory described by
\be
f=R+ \epsilon R^2 \; , \;\;\;\; \phi=1\;, \;\;\;\; \omega=1\;, \;\;\;\; V=0 \;.
\ee
General relativity is recovered by letting $\epsilon \rightarrow 0$. One has, for both theories,
\be
f_0=0 \;, \;\;\;\; F_0=1 \;,\;\;\;\; V_0=V_0'=f_0'=f_0''=f_{\phi R}=f_{\phi\phi}=f_{RR}=0 \;.
\ee
Eqs.~(\ref{64}) and (\ref{50}) yield $\Delta F=0$ and $\Phi_H=\Phi_A=0$. Eq.~(\ref{52}) then yields
$\Delta R=0$ and eq.~(\ref{1000}) becomes
\be \label{lilly}
\Delta\ddot{\phi}+k^2 \Delta \phi =0\;.
\ee
Eqs.~(\ref{lilly}) and (\ref{1002}) describe the evolution of scalar and tensor 
inhomogeneous perturbations of Minkowski space, which are effectively decoupled. 
It is obvious from these equations that Minkowski space is stable since the frequency squared 
is positive in each of these equations.

\vskip0.8truecm
Let us consider now the class of scalar-tensor gravity theories, which have 
$ f \left( \phi, R \right) =f(\phi) R $ and  $f_{RR}=0$. One  has 
to consider eqs.~(\ref{88}) and (\ref{89}), which become
\be
 \Delta R = -\frac{3 f_0'}{f_0} \left( \Delta \ddot{\phi} + k^2 \Delta \phi \right) \;,
\ee
\be  \label{524}
 \Delta \ddot{\phi}+  \left[ k^2 + \frac{ f_0 \left( 2V_0'' - f_0''\right)}{2f_0 +3f_0' } 
\right] \Delta \phi =0 \;.
\ee
The stability of a scalar perturbation depends on its wavelength and is achieved if 
\be 
k  \geq k_{14}= \left[ \frac{ f_0 \left( f_0'' - 2V_0'' \right) }{ 2f_0 +3f_0' } \right]^{1/2}
 \;,
\ee
when the argument of the square root is positive (or for any wavelength if the latter is negative or 
zero).

%%%%%%%%%%%%%%%%%%%%%%%%%%%%%%%%%%%%%%%%%%%%%%%%%%%%%%%%%%%%%%%%%%%%%%%%%%%%%%%%%%%%

\section{Discussion and conclusions}
\setcounter{equation}{0}

Motivated by inflation, quintessence, and quantum gravity corrections to the low-energy 
gravitational action,  we have derived conditions for the existence and linear 
stability of de Sitter 
solutions in a  very general theory, including scalar-tensor gravity, induced 
gravity, non-linear gravity, $1/R$ corrections to the Einstein-Hilbert action, non-minimally 
coupled scalar field theory, general relativity with or without a minimally coupled scalar 
and a cosmological constant, and phantom fields. Our analysis does not depend on the specific 
form of the coupling 
functions and the conditions for the existence and stability of de Sitter solutions are, in 
this 
respect, very  general. Minkowski space is studied as a special case of de Sitter space.
The phase space picture of the theory depends in an essential way on 
the form of the coupling functions and the scalar field potential $V( \phi )$ but, in a 
spatially flat FLRW universe, the dynamical variables are always $H$ and $\phi$. Although 
the field equations are of fourth order in non-linear theories of gravity, and 
hence 
the dimensionality of the phase space  depends crucially on the form of $f( \phi, R 
)$, the fixed points of the dynamical system are always de Sitter spaces with constant scalar 
field. It is for this reason that the conditions for the existence of de Sitter space and  for 
its stability can be expressed by inequalities valid for any choice of the coupling functions 
and the values of the free parameters.

Eqs.~(\ref{14}) and (\ref{15}) are necessary and sufficient conditions for the 
existence of de Sitter fixed points. Note that the existence of these de Sitter solutions is 
not automatically guaranteed in generalized gravity. Eqs.~(\ref{14}) and (\ref{15}) 
reduce to conditions  previously obtained in special cases of generalized gravity 
(non-minimally coupled scalar field theory \cite{FaraoniPLA,FaraoniIJTP} or  theories with $f 
=AR^n$ \cite{BarrowOttewill83}).

When de Sitter fixed points exist, their stability against linear inhomogeneous perturbations and 
their attractor behaviour are assessed by using a covariant and gauge-invariant 
formalism originally developed to study perturbations of FLRW spaces.  
It is established that expanding (resp. contracting) de Sitter spaces are always stable 
(resp. unstable) 
with respect to tensor perturbations. Scalar perturbations may threaten the stability of 
expanding 
de Sitter spaces, which are the ones of interest for inflationary and quintessence scenarios of 
the real universe. For non-linear theories of gravity (with $f_{RR} \neq 0$)
Subsection 3.2 provides the desired stability conditions, while 
section~5 describes the stability of Minkowski space. For linear gravity theories with 
$f_{RR}=0$, including scalar-tensor gravity and general relativity, the stability conditions 
for de Sitter space are given by eq.~(\ref{93}), while the stability of Minkowski space is 
determined by eq.~(\ref{524}). In the particular case of non-minimally coupled scalar field 
theory, these conditions reproduce those already known from a previous analysis 
\cite{FaraoniPLA,FaraoniIJTP}.

The analysis presented here can be generalized further. First, it is well known that there can 
be attractors in phase space that are inflationary but are not fixed points. This is the case 
of power-law  inflation $a(t)=a_0 \, t^p$ with $p>1$, which is an attractor solution of 
Brans-Dicke cosmology when the only form of matter present is a cosmological constant. 
Extended inflationary scenarios \cite{extended} are based 
on the presence of this attractor. Power-law inflation is also an attractor in scalar-tensor 
theories  generalizing Brans-Dicke gravity \cite{VF04} and in many non-linear theories 
\cite{Mulleretal90}. Hence, in general, the fixed points do not provide the complete 
phase space picture. Second, we restricted our attention to inhomogeneous perturbations. 
Although more general than the case of homogeneous perturbations usually studied in the 
literature, it would be interesting to generalize the stability analysis to anisotropic 
and to nonlinear perturbations. Finally, we considered only four spacetime dimensions but quantum 
gravity extensions of general relativity would call for a more general 
analysis in arbitrary spacetime dimension.

The stability conditions derived here can be applied to the investigation of the Big Rip 
singularity in the future. It has been pointed out that 
 the present expansion of the universe may be  superaccelerated, i.e., 
$ \dot{H}>0$, which is equivalent to an effective equation of state parameter $w \equiv 
P/\rho <-1$ for the dark energy dominating the dynamics of the universe at redshifts $z \leq 1$
 \cite{bigrip}. Superacceleration cannot be achieved with a canonical, minimally coupled scalar 
field in 
Einstein gravity \cite{VF04,bigrip}. If the universe really superaccelerates 
(which is not yet established due to the error in the observational determination of the 
parameter $w$) it 
runs the risk of ending in a Big Rip singularity in a finite future \cite{bigrip}. This 
kind of singularity is different from the Big Bang or the Big Crunch because the 
universe expands explosively while the energy density of dark energy diverges 
instead of getting diluted, due to the peculiar equation of state $P<-\rho$ 
\cite{VF04,bigrip}. If the equation of state of the dark energy is constant with 
$w=$const.$<-1$, the Big Rip is unavoidable. However, a time-dependent effective 
equation of state with $w=w(t)$ is more realistic and in this case scenarios have been proposed
 in which the Big Rip is avoided. At present, there are in the literature 
superaccelerating models 
in which the Big Rip is unavoidable \cite{bigrip} and others in which a late time de 
Sitter attractor with $\dot{H}=0$ exists which stops superacceleration and avoids the Big Rip 
\cite{Singhetal,SamiToporensky,HaoLi}. It is in this context that the stability conditions 
derived here can play a role: 
these conditions help assessing the stability of de Sitter spaces and 
deciding whether a late time de Sitter attractor exists that attracts the orbits of the 
solutions of the field equations in phase space, thus avoiding the Big Rip.
 A generic statement about the fate of the universe 
requires the knowledge of the attraction basin of an attractor, and this issue can only be 
addressed in a specific theory with the  form of the potential $V( \phi )$ fixed. However, the 
conditions for the existence of de Sitter attractors provide an answer about the possibility 
of avoiding the Big Rip at least for initial conditions lying in a certain attraction basin to 
be determined.

The stability conditions derived in this paper will be applied elsewhere to specific models of 
inflation and dark 
energy.

%%%%%%%%%%%%%%%%%%%%%%%%%%%%%%%%%%%%%%%%%%%%%%%%%%%%%%%%%%%%%%%%%%%%%%%%%%%%%%%%%%%%
\section*{Acknowledgments}

We acknowledge a referee for comments leading to improvements in the presentation of 
this material.  This work was supported by a Discovery Grant from 
the National Science and Engineering Research Council of Canada ({\em NSERC}).

%It is  a pleasure to thank    and  for 
%stimulating discussions.
%%%%%%%%%%%%%%%%%%%%%%%%%%%%%%%%%%%%%%%%%%%%%%%%%%%%%%%%%%%%%%%%%%%%%%%%

\clearpage

\section*{Appendix A: stability analysis for de Sitter space}

\def\theequation{A.\arabic{equation}}\setcounter{equation}{0}

The stability of the system (\ref{birba}) is assessed by studying the sign of  the real part of 
the eigenvalues $\lambda $ of the matrix $ \hat{M} $. The characteristic equation Det$\left( 
\hat{M} -\lambda \hat{I} \right)$, 
where $ \hat{I} $ is the  identity matrix, reduces to
\be  \label{chardS}
\lambda^4+B_1 \lambda^2 +C_1 = 0 \;,
\ee
where 
\begin{eqnarray}
B_1 & = & 2 k^2 + \frac{\alpha}{\eta_0^2} + \frac{\gamma}{\eta_0^2} \;,\\
&& \nonumber \\
C_1 &= &  \left(  k^2 + \frac{\alpha}{\eta_0^2} \right)  + \left( k^2 + \frac{\gamma}{\eta_0^2}\right) 
-\, \frac{ \beta \, \delta}{ \eta_0^4} \;.\\
\end{eqnarray}

The squares of the roots are given by 
\be
\lambda^2_{\pm} = \frac{ -B_1 \pm \sqrt{ \Delta_1}}{2} \;, \;\;\;\;\;\;\;\;\;\;
\Delta_1 = B_1^2-4C_1 \;.
\ee
Let us consider the  cases $ C_1 > 0 $, $ C_1 = 0 $, and $ C_1 < 0 $ separately. \\\\
\noindent $\bullet \;\; 0< C_1 \leq  B_1^2/4$ \\\\
Note that one cannot have $ B_1 = 0 $ in this case. If $ B_1 > 0 $, then $\lambda_{ \pm }^2 <0$ and 
the roots 
$\lambda_{\pm\,\pm} $ of eq.~(\ref{chardS}) are purely  imaginary, i.e., $Re\left( \lambda \right)=0$ 
and de Sitter  space is neutrally stable.

If $B_1 < 0 $ then $\lambda_{\pm}^2 >0$ and all the four eigenvalues $\lambda_{\pm\,\pm} $ are real, 
with two of them positive and two negative. The positive ones give rise to instability.\\\\
\noindent $\bullet \;\;  C_1 > B_1^2/4  $ ~~~~($ \Delta_1 <0$) \\\\
If $B_1 \neq 0$, then 
\be
\lambda_{\pm}^2 =\frac{ -B_1 \pm i \sqrt{ \left| \Delta_1 \right|}}{2} \equiv \rho \, 
\mbox{e}^{i\theta_{\pm}} 
\ee
and
\be
\lambda_{\pm \, \pm} = \pm  \sqrt{ \rho} \, \mbox{e}^{i \, \frac{\theta_{\pm}}{2}  } \; .
\ee
In this case two roots $\lambda $ have positive real part and de Sitter space 
is unstable.

If $ B_1 = 0 $ and $ C_1 > 0$ one has $\lambda^2_{\pm}=\pm i \sqrt{  C_1 }$ and
\be
\lambda_{\pm\pm}=\pm \frac{ C_1^{1/4}}{\sqrt{2}} \left( 1\pm i \right) \;.
\ee
The roots with positive real part are associated with instability of de Sitter space.\\\\
\noindent $\bullet \;\;  C_1=0   $\\\\
In this case eq.~(\ref{chardS}) gives $\lambda_1^2 =0$ or $\lambda_{2,3}^2=-B_1 $. If $B_1 > 0$ the 
roots 
$ \lambda_{2,3} $ are 
purely imaginary, corresponding to oscillating perturbations and to stability. If instead $ B_1 < 0 $ 
there is a real positive root associated with instability. If $ B_1 = C_1 = 0 $ then $\lambda=0$ and de 
Sitter  space is neutrally stable.\\\\
\noindent $\bullet \;\;  C_1 < 0 $\\\\
In this case
\be
\lambda^2=\frac{ -B_1 \pm \sqrt{ B_1^2+4|C_1|}}{2} \;.
\ee
The lower sign produces two imaginary roots, while the upper sign gives two real positive roots 
associated with instability.

As a summary, de Sitter space is stable if  
\be
0 \leq C_1 \leq \frac{B_1^2}{4} \;\;\;\;\;\;\;\;\; \mbox{and} \;\;\;\;\;\;\;\;\; B_1 \geq 0 
\ee
and unstable otherwise.

The conditions for stability can be formulated in terms of effective Jeans wavelengths and of 
the values of the coupling functions and parameters of the theory. The inequality $C_1 \geq 0$ is 
equivalent to 
\be
2k^2 \eta_0^2 + \alpha + \gamma -\beta \delta \geq 0 \;,
\ee
which can be expressed as
\be
k \geq k_1 \equiv \left\{ 
\frac{1}{2\eta_0^2} \left[ \frac{1}{2\omega_0} \left( f_0''-2V_0'' -\,\frac{ f_{\phi R}^2}{ f_{RR}} 
+ \frac{ F_0 f_{\phi R}^2}{ 3H_0^4 f_{RR}^2 } \right) +8 -\,\frac{F_0}{3f_{RR}H_0^2} 
\right] \right\}^{1/2} 
\;.
\ee
The inequality $B_1 \geq 0$ is equivalent to 
\be
2k^2 \eta_0^2 + \alpha + \gamma \geq 0 \;,
\ee
or
\be \label{bizz}
b_1 + 2 + k^2 \geq 0 \;,
\ee
where
\be
b_1 =  \frac{1}{ \eta_0^2} \left[ \frac{1}{2\omega_0} \left( 2V_0''-f_0''
+\frac{ f_{\phi R}^2}{f_{RR}} \right) + \frac{ F_0 }{3 f_{RR} H_0^2} -8  \right] -2
\;.
\ee
Eq.~(\ref{bizz})  is satisfied for any $k$ if $b_1 +2 \geq 0$ or, if $b_1 < -2 $, by wave 
vectors $k$ such that
\be
k \geq k_2 \equiv \left\{ 
\frac{1}{2\eta_0^2} \left[ \frac{1}{2\omega_0} \left( f_0''-2V_0'' -\,\frac{ f_{\phi R}^2}{ f_{RR}} 
\right) + 8 - \frac{ F_0 }{ 3H_0^2 f_{RR} }  \right] \right\}^{1/2} =\sqrt{ \left| 
b_1+2\right| } \;.
\ee
The inequality $ C_1 \leq B_1^2/4$ is equivalent to
\be
k^4 + \left[ \frac{ \alpha+\gamma}{\eta_0^2} -2 \right] k^2 
+ \, \frac{ \left( \alpha+\gamma \right)^2}{4\eta_0^4} 
- \,\frac{ \left( \alpha+\gamma \right)}{\eta_0^2}
+ \frac{ \beta \, \delta }{ \eta_0^4} \geq 0 \;, 
\ee
which can be written as
\be \label{buzz}
\varphi( k) \equiv k^4+b_1 \, k^2 +c_1 \geq 0 \;,
\ee
where
\begin{eqnarray}
c_1 & = & \frac{1}{ 4 \eta_0^4} \left[ 
\frac{1}{2\omega_0} \left( 2V_0''-f_0''+\frac{ f_{\phi R}^2}{f_{RR}} \right) -8 
+\frac{ F_0}{3f_{RR} H_0^2} \right]^2  \nonumber \\
&& \nonumber \\
&&   - \frac{1}{\eta_0^2}  \left[ \frac{ F_0}{3f_{RR}H_0^2} 
+ \frac{1}{2\omega_0} \left( 2V_0'' -f_0'' + \frac{ f_{\phi R}^2 }{f_{RR}} \right) -8 \right] +\, 
\frac{ F_0 f_{\phi R}^2 }{6 \omega_0 f_{RR}^2 H_0^4 \eta_0^4 }  \;.
\end{eqnarray}
To identify the values of the wave vector $k$ that satisfy the inequality (\ref{buzz}) one 
studies the sign of $\varphi (k)$ by distinguishing several cases.

\begin{itemize}

\item $b_1 \geq 0$ and $c_1 \geq 0$: the inequality (\ref{buzz}) is satisfied for any value of $k$.

\item $b_1 \geq 0$ and $c_1 < 0$: the curve representing $\varphi(k)$ starts negative at $k=0$ and is 
always increasing, crossing the $k$-axis at a point $k_3$. The function $\varphi(k)$  becomes 
positive  for $k> k_3$, where 
\be
k_3= \sqrt{ \frac{ -b_1 + \sqrt{ b_1^2-4c_1}}{2} } \;.
\ee

\item $b_1 =0 $: then, if $c_1 \geq 0$,  $ \varphi(k) >0 $ for any value of $k \geq 0$;\\ 
if $c_1  < 0$ then $ \varphi(k) \geq 0 $ for  $ k \geq  k_4 = \left| c_1 \right|^{1/4}$.

\item $b_1 < 0$: the curve representing $\varphi(k) $ starts from the value $c_1 $ at $k=0$, decreases 
for $0 < k < \sqrt{ \left| b_1 \right|/2} $, reaches a minimum at $
k=\sqrt{ \left| b_1 \right|/2}$, and then is always increasing for $ k> \sqrt{ \left| b_1 
\right|/2}$.
If $ c_1> 0$ and the minimum is non-negative, then $\varphi(k) \geq 0$ for any value of $k$ and the 
equation $\varphi( k) =0 $ has no real roots. This happens if $c_1 > b_1^4/4$.\\
If instead $c_1 >0$ but the minimum of $\varphi(k) $ is negative there are two real roots of the 
equation $\varphi(k)=0$ and $ \varphi  $ is positive for 
$ 0< k< k_5 $ and for $k> k_6$, and negative otherwise. This situation occurs if $ c_1 \leq 
b_1^2/4$ and the critical wave vectors are 
\be
k_{5,6} = \sqrt{ \frac{  \left| b_1 \right| \pm \sqrt{ b_1^2-4c_1}  }{2} } \;.
\ee
Finally, if $ c_1<0$, the curve representing $\varphi(k) $ is negative for $ 0 \leq k < k_7 $ and 
positive for 
\be
k> k_7 =
 \sqrt{ \frac{  \left| b_1 \right| + \sqrt{ b_1^2 + 4 \left| c_1 \right| }  }{2} } \;.
\ee
\end{itemize}

By putting together the three conditions for stability $C \geq 0$, $B \geq 0$ and $ C \leq B^2/4$ one 
obtains that, in order for de Sitter space to be stable with respect to scalar perturbations, the 
latter must have wave vectors in one of the  following intervals:

\begin{itemize} 

\item if $ b_1 \geq 0$ and $c_1 \geq 0$, it must be $ k\geq k_1 $ for 
stability

\item if $ b_1 > 0$ and $c_1 < 0$, it must be $ k\geq \mbox{max}\left\{ k_1, k_3 \right\} $

\item if $ b_1 = 0$ and $c_1 \geq 0$, it must be $ k\geq  k_1 $

\item if $ b_1 = 0$ and $c_1 < 0$, it must be $ k\geq \mbox{max}\left\{ k_1, k_4\right\} $

\item if $ -2 \leq  b_1 < 0$ and $c_1 > b_1^2/4 $, it must be $ k\geq  k_1 $

\item if $ -2 \leq b_1 < 0$ and $ 0 < c_1 < b_1^2/4  $, it must be 
$  k_1 \leq k \leq k_5 $ (note that this inequality can be satisfied only if $k_5 > k_1, k_2 $) or
 $ k \geq  \mbox{max}\left\{ k_1, k_6 \right\} $

\item if $ -2 \leq b_1 <  0$ and $c_1 < 0$, it must be $ k \geq \mbox{max}\left\{ k_1,  
k_7 \right\} $

\item if $  b_1 <  -2 $ and $c_1 < 0$, it must be $ k\geq \mbox{max}\left\{ k_1, k_2, k_7  
\right\} $

\item if $ b_1 < -2 $ and $ 0< c_1 \leq b_1^2 /4  $, it must be $ \mbox{max} 
\left\{ k_1, k_2 \right\} \leq k \leq k_7 $ or  $ k \geq \mbox{max}\left\{ k_1, k_2, k_6  
\right\} $

\item if $ b_1 <  -2 $ and $c_1 > b_1^2/4 $, it must be $ k \geq \mbox{max}\left\{ k_1, 
k_2 \right\} $.

\end{itemize} 

%%%%%%%%%%%%%%%%%%%%%%%%%%%%%%%%%%%%%%%%
\section*{Appendix B: stability analysis for Minkowski space}

\def\theequation{B.\arabic{equation}}\setcounter{equation}{0}

The stability of the system (\ref{1007}) is determined by the real part of the 
eigenvalues $\lambda $ of $ \hat{N} $. The characteristic equation Det$\left( \hat{N} -\lambda 
\hat{I} 
\right)$ is
\be  \label{charM}
\lambda^4+B_2\lambda^2 + C_2 =0 \;,
\ee
where 
\begin{eqnarray}
B_2 & = & k^2 + \frac{F_0 \, f_{\phi R} }{ 3f_{RR} } -D \;,\\
&& \nonumber \\
C_2 &= &  - D k^2 -\, \frac{F_0 \, f_{\phi R} }{ 3f_{RR} } \left( D+    
\frac{ f_{\phi R} }{ 2\omega_0 \, f_{RR} } \right) \;.
\end{eqnarray}

The squares of the roots are given by 
\be
\lambda^2_{\pm} = \frac{ -B_2 \pm \sqrt{ \Delta_2}}{2} \;, \;\;\;\;\;\;\;
\Delta_2 = B_2^2-4C_2 \;.
\ee
Let us consider the  cases $ C_2 > 0 $, $ C_2 = 0 $ and $ C_2 < 0 $ separately. \\\\
\noindent $\bullet \;\; 0< C_2 \leq  B_2^2/4$ \\\\
Note that one cannot have $ B_2 = 0 $ in this case. If $ B_2 > 0 $, then $\lambda_{ \pm }^2 <0$,  
the roots $\lambda_{\pm \, \pm} $ of eq.~(\ref{charM}) are purely  imaginary, and 
Minkowski space is neutrally stable.

If $ B_2 < 0 $ then $\lambda_{\pm}^2 >0$ and all the four  eigenvalues are real, with two of them  
positive and two negative -- the positive ones make Minkowski space unstable.\\\\
\noindent $\bullet \;\;  C_2 > B_2^2/4  $\\\\
If $ B_2 \neq 0$, then 
\be
\lambda_{\pm}^2 =\frac{ -B_2 \pm i \sqrt{ \left| \Delta_2 \right|}}{2} \equiv \rho \, 
\mbox{e}^{i\theta_{\pm}} 
\ee
and
\be
\lambda_{\pm \, \pm} = \pm  \sqrt{ \rho} \, 
\mbox{e}^{i \, \frac{\theta_{\pm}}{2}  } \; . 
\ee
In this case two roots $\lambda_{ \pm \, \pm} $  have positive real part and 
Minkowski space is unstable.

If $ B_2 = 0 $ and $ C_2 > 0 $ one has 
$\lambda_{\pm \, \pm} = \pm 2^{-1/2} C^{1/4} \left( 1 \pm i \right) $. Two roots have positive real 
part, corresponding to instability.\\\\
\noindent $\bullet \;\;  C_2 = 0  $ \\\\
In this case eq.~(\ref{charM}) gives $\lambda^2 =0$ or $\lambda^2=-B_2 $. If $ B_2 > 0$ the roots are 
purely imaginary, corresponding to oscillating perturbations and to stability. If instead $ B_2 < 0 $ 
there 
is a real positive root associated with instability. If $ B_2=C_2=0 $ then $\lambda=0$ and Minkowski 
space is neutrally stable.\\\\
\noindent $\bullet \;\;  C_2 < 0 $\\\\
In this case
\be
\lambda^2=\frac{ -B_2 \pm \sqrt{ B_2^2+4|C_2|}}{2} \;.
\ee
The lower sign produces two imaginary roots, while the upper sign gives two real roots, one of which 
is positive and is associated with instability.

To summarize, Minkowski space is stable if  
\be
0 \leq C_2 \leq \frac{B_2^2}{4} \;\;\;\;\;\;\;\;\; \mbox{and} \;\;\;\;\;\;\;\;\; B_2 \geq 0 
\ee
and unstable otherwise.

Let us express  the conditions above in terms of the coupling functions  and of the 
wave vector $k$. The inequality $ C_2 \geq 0$ is 
equivalent to
\be    \label{ineq1}
\psi (k) \equiv k^4 +b_2 k^2 +c_2 \geq 0 \;,
\ee
where 
\begin{eqnarray}
b_2 & = & \frac{1}{2\omega_0} \left( 2V_0'' -f_0'' +\frac{ f_{\phi R}}{f_{RR}} \right)
+ \frac{ F_0 f_{\phi R} }{ 3f_{RR} } \;, \\
&& \nonumber \\
c_2 & = &  \frac{F_0 f_{\phi R}}{6\omega_0 f_{RR}}  \left( 
2V_0'' -f_0''  \right)  \;.
\end{eqnarray}
The inequality $ B_2 \geq 0$ is equivalent to 
\be  \label{ineq2}
k^2 + \frac{b_2}{2} \geq 0 \;,
\ee
which is always satisfied if $b_2 \geq 0$, and is satisfied only for perturbations with wave vectors 
such that
\be 
k \geq k_8 =\sqrt{ \frac{ |b_2|}{2}} 
\ee
when $b_2<0$. The inequality $C_2 \leq  B_2^2 / 4 $ is equivalent to 
\be         \label{ineq3}
4c_2 \leq b_2^2 \;.
\ee
Note that in this case the wave vector drops out and this is a requirement on the theory of gravity  
independent of the wavelength of the inhomogeneous perturbation. Next, one studies the sign of 
$\psi(k) $ as in Appendix~A, with the following result.

\begin{itemize}

\item $b_2 \geq 0$ and $c_2 \geq 0$: then $\psi(k) \geq 0$ for any value of $k$.

\item $b_2 \geq 0$ and $c_2 < 0$: it is  $\psi (k)\geq 0 $ for 
\be
k \geq  k_9=  \sqrt{ \frac{ -b_2 + \sqrt{ b_2^2-4c_2}}{2} } \;.
\ee

\item $b_2 =0 $: if $c_2 \geq 0$ then $ \psi(k) >0 $ for any value of $k$;\\ 
if $c_2  < 0$ then $ \psi(k) \geq 0 $ for  $ k \geq  k_{10} = \left| c_2 \right|^{1/4}$.

\item $b_2 < 0$: if  $ c_2 > b_2^2/4$, then $\psi(k)>0$ for any $k$.\\
If $ 0<c_2 \leq  b_2^2/4$, then $\psi(k) >0$ for 
$0 < k < k_{11}$ and for $k>k_{12}$, where 
\be
k_{11,12} = \sqrt{ \frac{  \left| b_2 \right| \pm \sqrt{ b_2^2-4c_2}  }{2} } \;.
\ee
Finally, if $ c_2 < 0 $,  $\psi(k) $ is positive for 
\be
k > k_{13} =
 \sqrt{ \frac{  \left| b_2 \right| + \sqrt{ b_2^2 + 4 \left| c_2 \right| }  }{2} } \;.
\ee
\end{itemize}

The stability of Minkowski space is assured by imposing that the three inequalities (\ref{ineq1}), 
(\ref{ineq2}) and (\ref{ineq3}) hold simultaneously. It must be 
$ 4c_2 \leq b_2^2$,  plus one of the following conditions must hold:

\begin{itemize} 

\item $ b_2 = 0$, $c_2 \leq 0$, and $k\geq \mbox{max} \left\{ k_8, k_{10} \right\} $

\item $ b_2 < 0$, $ c_2 < 0$, and $ k\geq \mbox{max}\left\{ k_8,  k_{13} \right\} $

\item $ b_2 > 0$ and  $   0\leq c_2 \leq  b_2^2/4 $

\item $ b_2 > 0$, $c_2 < 0$, and  $ k\geq k_9  $.

\end{itemize}

%%%%%%%%%%%%%%%%%%%%%%%%%%%%%%%%%%%%%%%%%%%%%%%%%%%%%%%%%%%%%%%%%%%%%%%%
   
\end{document}